\documentclass{IEEEtran}
\usepackage{graphicx}
\usepackage[T1]{fontenc}% optional T1 font encoding
\usepackage{amsmath,amsfonts,amssymb}
\usepackage{bm}
\usepackage{algorithmic}
\usepackage{array}
\usepackage{comment}
\usepackage{stfloats}
\usepackage[usenames,dvipsnames]{xcolor}
\usepackage{color,url,bbding,soul}
\usepackage{gensymb}
\usepackage[caption=false,font=footnotesize]{subfig}
\usepackage[export]{adjustbox}
\usepackage{multirow}
\usepackage{enumitem}
\usepackage{dashbox}
\usepackage[linesnumbered,ruled,commentsnumbered]{algorithm2e}
\usepackage{tabularx}
\usepackage{wasysym}
\usepackage{etoolbox}
\makeatletter
\def\do#1{\patchcmd{#1}{\thepage}{\null}{}{\GenericWarning{}{Could not patch \string#1}}}
\docsvlist{\@oddhead,\@evenhead,\ps@headings,\ps@IEEEtitlepagestyle,\ps@IEEEpeerreviewcoverpagestyle}
\makeatother
\newcommand{\ignore}[2]{\hspace{0in}#2}
\newcolumntype{P}[1]{>{\centering\arraybackslash}p{#1}}
\bstctlcite{ICINBSTcontrol}

\def\house{\hbox{\kern3pt \vbox to13pt{}% 
		\pdfliteral{q 0 0 m 0 5 l 5 10 l 10 5 l 10 0 l 7 0 l 7 5 l 3 5 l 3 0 l f
			1 j 1 J -2 5 m 5 12 l 12 5 l S Q }%
		\kern 13pt}}

\begin{document}
\title{Beyond WYSIWYG: Sharing Contextual Sensing Data Through mmWave V2V Communications}	
\author{
	\IEEEauthorblockN{Cristina Perfecto\IEEEauthorrefmark{1}, Javier Del Ser\IEEEauthorrefmark{1}\IEEEauthorrefmark{2}\IEEEauthorrefmark{3}, Mehdi Bennis\IEEEauthorrefmark{4}\IEEEauthorrefmark{5} and Miren Nekane Bilbao\IEEEauthorrefmark{1}}\\
 
\IEEEauthorblockA{\hspace{-.17cm}\IEEEauthorrefmark{1}University of the Basque Country UPV/EHU, Spain, emails:\{cristina.perfecto,javier.delser,nekane.bilbao\}@ehu.eus}\\
\IEEEauthorblockA{\IEEEauthorrefmark{2}TECNALIA, Spain, email: javier.delser@tecnalia.com}\\
\IEEEauthorblockA{\IEEEauthorrefmark{3}Basque Center for Applied Mathematics (BCAM), Spain}\\
\IEEEauthorblockA{\IEEEauthorrefmark{4}Centre for Wireless Communications, University of Oulu, Finland, email: bennis@ee.oulu.fi}\\
\IEEEauthorblockA{\IEEEauthorrefmark{5}Department of Computer Engineering, Kyung Hee University, South Korea}
}
% make the title area
\maketitle
	\begin{abstract}
	In vehicular scenarios context awareness is a key enabler for road safety. However, the amount of contextual information that can be collected by a vehicle is stringently limited by the sensor technology itself (e.g., line-of-sight, coverage, weather robustness) and by the low bandwidths offered by current wireless vehicular technologies such as DSRC/802.11p. Motivated by the upsurge of research around millimeter-wave vehicle-to-anything (V2X) communications, in this work we propose a distributed vehicle-to-vehicle (V2V) association scheme that considers a quantitative measure of the potential value of the shared contextual information in the pairing process. First, we properly define the utility function of every vehicle balancing classical channel state and queuing state information (CSI/QSI) with context information i.e., sensing content resolution, timeliness and enhanced range of the sensing. Next we solve the problem via a distributed many-to-one matching game in a junction scenario with realistic vehicular mobility traces. It is shown that when receivers are able to leverage information from different sources, the average volume of collected extended sensed information under our proposed scheme is up to 71\% more than that of distance and minimum delay-based matching baselines.
\end{abstract}

\IEEEpeerreviewmaketitle
%\IEEEpubid{978--1--5386--3873--6/17/\$31.00~\copyright~2017 IEEE }

%%%%%%%%%%%%%%%%%%%%%%%%%%%%%%%%%%%%%%%%%%%%%%%%%%%%%%%%%%%%%%%%%%%%%%%%%%%%%%%%%%
\section{Introduction}

Autonomous vehicles equipped with devices for data sensing and communication lie at the cornerstone where the so-called Internet of Things (IoT) and cloud computing paradigms meet each other \cite{gerla2014internet}. From a macroscopic point of view, the construction of autonomous navigation routes relies on global positioning (e.g., GPS) and map services updated with information about current road conditions, which is provided by cloud computing. However, autonomous vehicles will need to determine their real-time moving strategy and decisions depending on the dynamic surroundings, for which sensors installed on board are of utmost necessity, especially --yet not uniquely-- for safety reasons. Indeed, a growing body of literature has emphasized the crucial role of vehicular sensors for the pre-collision detection of obstacles \cite{park2016survey} and platooning, \ignore{\cite{jia2016survey}} among many other applications \cite{vahdat2016survey,broggi2016intelligent}.
%\IEEEpubidadjcol
In this context, a vehicular sensor system is usually composed by light detection and ranging (LiDAR), radar, GPS, odometry, and computer vision devices\ignore{\cite{baber2005cooperative}}. Among this portfolio of sensing technologies, LiDAR has garnered the attention of car manufacturers due to the significantly higher spatial resolution provided by this radar technology, to the point of having being deemed the ``eyes'' of driverless vehicles \cite{rasshofer2005automotive}. For instance, a typical commercial LiDAR using 64 laser diodes produces 2.8 million data points per second with a $360^\circ$ horizontal field of view, a $26.8^\circ$ vertical field of view and a coverage range of more than 70m in all directions, generating a precise three-dimensional map of a car's surroundings. Sharing a small fraction of this sensing requires massive data rates. \ignore{By virtue of LiDAR, vehicles can detect obstacles and build 3D surroundings for safe navigation in dynamic environments.} 

However, the increased contextual awareness enabled by a vehicular sensor is restricted by its covered range centered on the position of the vehicle itself \cite{sattar2016recent}. To overcome this restriction, vehicular communications have been identified as a technology enabler to compensate for the shortcomings of sensors, break the line-of-sight constraint, acquire more data on
surroundings (e.g., blind area information), and ultimately enhance the overall contextual awareness of connected vehicles, in terms of the geographical spread and/or quality of the information objects representing the monitored environment (contents) \cite{Chiti2016}, \cite{Fanaei2014}. As a result, autonomous vehicles can acquire more valuable traffic data to optimize their driving behavior and increase their safety levels \cite{santa2009sharing}.
%\IEEEpubidadjcol
Unfortunately, conventional wireless technologies for vehicular communications have very limited bandwidth. For example, the maximum bit rate supported by Dedicated Short-Range Communications (DSRC) is 27 Mb/s \cite{kenney2011dedicated}. On the contrary, traffic sensor data are ever growing, supporting not only LiDAR's 3D imaging but also incorporating high-definition cameras. In this regard, the next-generation millimeter-wave (mmWave) wireless technology has been postulated as a feasible radio solution to solve this dilemma \cite{kong2017millimeter,Choi2016}\ignore{,Va2016a}.

mmWaves offer the potential to relax the ``what you \textit{sense} is what you get'' (WYSIWYG) rule by forming a swarm of vehicles that, connected through multiple vehicle-to-vehicle (V2V) links, expands each vehicle's own-sensing with real-time information retrieved from nearby vehicles.
%\hl{mmWaves have the potential to allow sharing real-time sensing data, forming a swarm of vehicles where  connected through multiple vehicle-to-vehicle (V2V) links information from nearby vehicles can be retrieved. Ultimately, each vehicle's own-sensing can be further enhanced and the ``what you \textit{sense} is what you get'' (WYSIWYG) rule relaxed}.
Indeed, such use of V2V mmWave links to enable \textit{connected intelligence} can effectively help to address both blind area and bad weather problems inherent to LiDAR and other sensing equipments. When a vehicle detects a blind area --as might happen when about to reach a junction--, sensor data from neighboring vehicles can be requested to augment situational awareness beyond its own sensing range. Besides, although the LiDAR range is dramatically reduced in bad weather, under these conditions the degradation undergone by mmWaves' transmission range is almost negligible. Therefore, mmWave communication counteracts the negative effects of a reduced sensing range. Concurrent mmWave transmissions --e.g., through beamforming or through multiple transceivers-- can be leveraged to simultaneously collect shared sensor data from several surrounding vehicles. Subsequently, 3D road condition reconstruction by multi-source multi-modal data analysis \cite{qian2016multi} can be performed. 

This paper builds upon the vibrant area of research of Cooperative Automated Driving (CAD) by proposing a distributed multi-beam association scheme for mmWave vehicular scenarios driven by the contextual value of the information shared among vehicles and delay constraints. The main idea is to expand the individual sensing range of vehicles by dynamically adding traffic/driving perception data from well-chosen neighbors to improve safety and traffic efficiency. The proposed method hinges on a set of novel utility functions for every vehicular transmitter (vTx) and vehicular receiver (vRx), based on which a distributed matching game is performed to find a stable many-to-one V2V association. Simulation results for an urban junction scenario with realistic mobility traces demonstrate that the proposed method increases the average amount of shared \emph{timely and innovative} contextual information when using narrow beams by up to 67\% (one-to-one matching) and 71\% (many-to-one) with respect to distance- or delay-based policies.

The rest of the paper is structured as follows: Section \ref{sec:system_model_and_problem} describes the system model and formulates the optimization problem tackled in this work. Next, Section \ref{sec:scheme} describes the many-to-one matching game to solve vTx and vRx pairing.\ignore{, taking into account interest and quality of offered information as well as sensing range extension among others} Simulation setup is presented and numerical results are discussed in Section \ref{sec:results}. Section \ref{sec:conclusions} ends the paper by sketching future research lines.

%%%%%%%%%%%%%%%%%%%%%%%%%%%%%%%%%%%%%%%%%%%%%%%%%%%%%%%%%%%%%%%%%%%%%%%%%%%%%%%%%%
\section{System Model and Problem Formulation}\label{sec:system_model_and_problem}
%%%%%%%%%%%%%%
\subsection{System Model} \label{sec:system_mode}
\begin{figure}[h!]
	\centering
	\includegraphics[width=1.03\columnwidth]{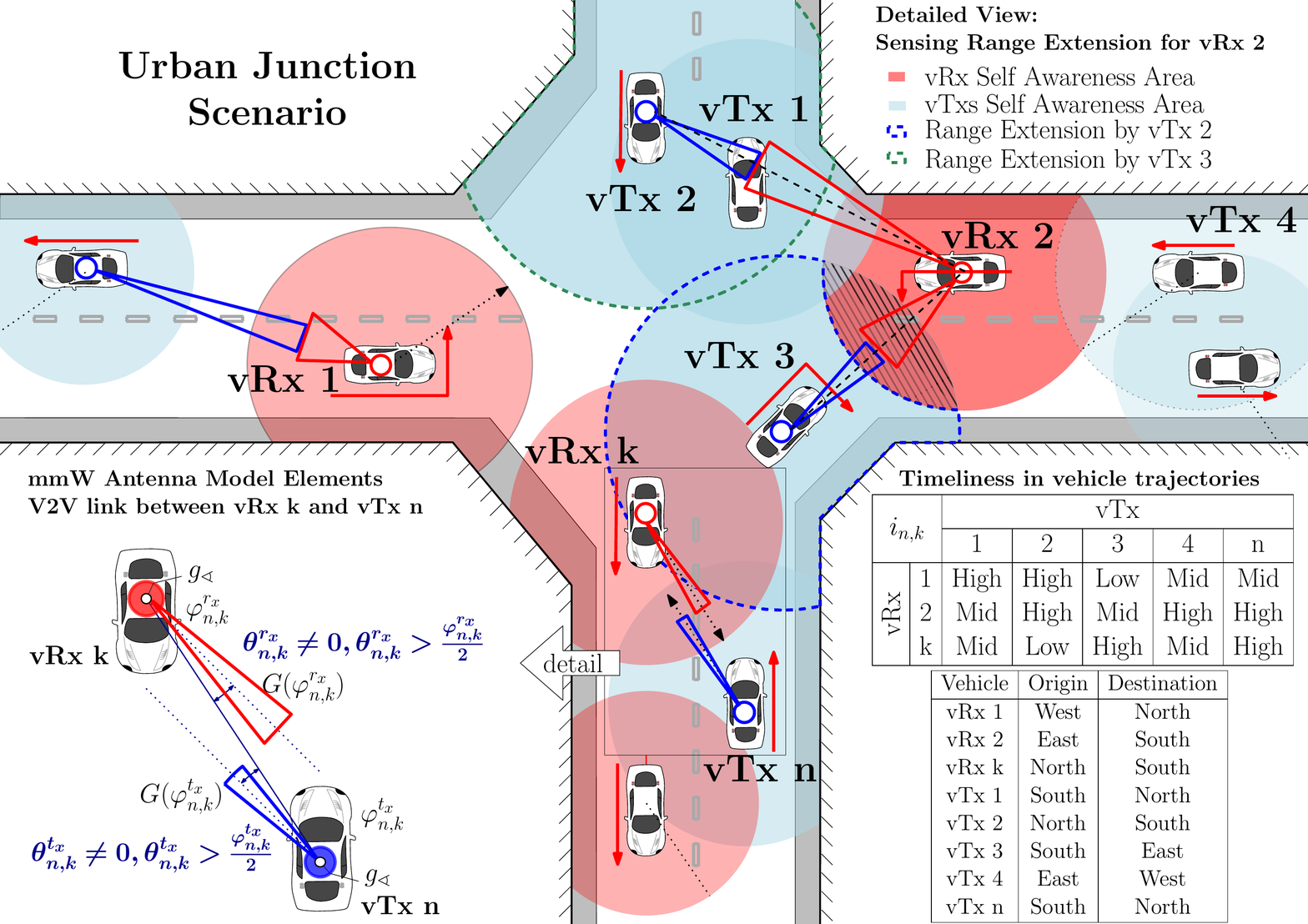}
	\caption{Snapshot of an urban junction with several transmitters (vTx) and receivers (vRx). vTx 1 and vTx 3 are connected to vRx 2 via mmWave links to increase the context awareness level of the latter depending on their route (marked with $\color{red} \bm{\rightarrow}$) and the extended area from which contextual data can be sensed and sent by the transmitter (likewise $\color{blue} \mbox{\textbf{- - -}}$ for vTx 3, $\color{OliveGreen} \mbox{\textbf{- - -}}$ for vTx 2).\vspace{-3mm}}
	\label{fig:1}
\end{figure}

An urban traffic junction scenario, as depicted in Fig. \ref{fig:1}, is considered. It is assumed that vehicles communicate through half-duplex V2V links over the 60 GHz mmWave frequency band with bandwidth $B$ and uniform transmit power $P$. Let $\mathcal{N}\triangleq\{1,\ldots,N\}$, $\mathcal{K}\triangleq\{1,\ldots,K\}$ respectively denote the sets of vTx and vRx. Without loss of generality and for sake of simplicity in foregoing discussions, time-slotted communications with transmission slots of duration $T_t$ seconds are adopted, whereas resource scheduling --which includes pairing and beam alignment between vTx and vRx-- is performed every $T_s$ seconds\ignore{, with $T_s/T_t$ simplified to be an integer number $M$}. Therefore, each scheduling slot will span $M$ transmission slots --i.e., $T_s=MT_t$--, such that scheduling is held at $\mathcal{T}_s\triangleq\{t_s\in \mathbb{N}: t_s\mod{M}=0\}$ and data transmission is held at $\mathcal{T}_t \triangleq \mathbb{N}$.

A log-distance pathloss model \cite{Yamamoto2008}, \hspace{-.5mm}\cite{HeathSurvey2016} is adopted that, through values of pathloss exponent $\vartheta_{n,k}$ and $\beta_{n,k}$ captures the impact of the amount of blocking vehicles obstructing the mmWave V2V link at hand. Under this model the channel gain $g_{n,k}^{c}$ for the link of $s_{n,k}$ meters from vTx $n$ to vRx $k$ is
\begin{equation}\label{eq:pathloss}
g_{n,k}^{c}\hspace{-.4mm}(t)\hspace{-1mm}=\hspace{-1mm}10\hspace{-.4mm}\:\vartheta_{n,k}(t)\hspace{-.8mm}\log_{10}(s_{n,k}(t))\hspace{-.8mm}+\hspace{-.8mm}\beta_{n,k}(t)\hspace{-.8mm}+\hspace{-.8mm}15\:\hspace{-.5mm}s_{n,k}(t)\hspace{-.2mm}/\hspace{-.2mm}1000.
\end{equation}
In the above channel gain, vehicles' movement entangles a dynamic amount of blockers in the link. Consequently, not only distance $s_{n,k}$, but also parameters $\vartheta_{n,k}$ and $\beta_{n,k}$ are time dependent.

Furthermore, on-vehicle mmWave directional antennas are modeled through a two-dimensional ideal sectored antenna \ignore{approximated by resorting to a parametric two-dimensional model that allows defining the boresight direction, the half-power beamwidth and front-to-back ratio for directivity gains in the mainlobe and sidelobe} whose transmission and reception antenna gains for vehicles over link $(n,k)$ during a transmission slot $t\in\mathcal{T}_t$, correspondingly $g_{n,k}^{t_x}\hspace{-.2mm}(t)$ and $g_{n,k}^{r_x}\hspace{-.2mm}(t)$ with $\wp\hspace{-.3mm}\in\hspace{-.7mm}\{t_x,r_x\}$, are given by \cite{Wildman:14} as
\begin{equation}\label{eq:antenna}
g_{n,k}^\wp(t)\hspace{-1.1mm}=\hspace{-1.1mm}\left\lbrace
\begin{array}{ll}
\hspace{-2.2mm}G\hspace{-0.3mm}(\varphi_{n,k}^\wp)\hspace{-1mm}=\hspace{-1mm}\frac{2\pi\mbox{-}\left(2\pi\mbox{-}\varphi_{n,k}^\wp\right)g_\sphericalangle}{\varphi_{n,k}^\wp}\text{,}\hspace{-1mm}&\hspace{-3mm}\text{if }\hspace{-0.5mm} |\theta_{n,k}^\wp(t)|\hspace{-1mm}\leq\hspace{-1mm}\frac{\varphi_{n,k}^\wp}{2}\\
\vspace*{-0.1cm}
\hspace{-1.4mm}g_\sphericalangle,\hspace{-1mm}&\hspace{-3mm}\text{otherwise.}\\
\end{array}	\right.
\end{equation}
In \eqref{eq:antenna} $\theta_{n,k}^\wp(t)$ stands for the alignment error between the steering and boresight directions of vTx $n$ and vRx $k$, and $\varphi_{n,k}^\wp$ is the mainlobe beamwidth of link $(n,k)$ at transmission ($\wp=t_x$) and reception ($\wp=r_x$) sides established for the active scheduling period. The non-negligible sidelobe power is given by $0\leq g_\sphericalangle\ll 1$. In order to perform beam alignment, a two-stage mechanism will be considered based on the trial-and-error procedure simplified from \cite{Wang2009}. The alignment delay $\tau_{n,k}$ under this procedure is given --on the assumption of availability of a priori sector knowledge provided by the sensing equipment \cite{Choi2016}-- by $\tau_{n,k}=\psi_n^{t_x}\psi_k^{r_x}T_p/(\varphi_{n,k}^{t_x}\varphi_{n,k}^{r_x})$, where $\psi_n^{t_x}$ and $\psi_k^{r_x}$ denote the sector-level beamwidths of vTx $n$ and vRx $k$, and $T_p$ is the pilot transmission duration. The rate for a time slot $t$ of duration $T_t$ within which alignment has been performed is
\begin{equation} \label{eq:rate}
r_{n,k}(t)=\left(1-\tau_{n,k}/{T_t}\right)B\log_2\left(1+\gamma_{k}(t)\right),
%r_{n,k}(t)=\left(1-\frac{\tau_{n,k}}{T_t}\right)B\log_2\left(1+\gamma_{k}(t)\right),
\end{equation}
\begin{equation} \label{eq:sinr}
\gamma_{k}(t)=\frac{p_n g_{n,k}^{t_x}(t)g_{n,k}^c(t)g_{n,k}^{r_x}(t)}{\sum_{\substack{n'\in\mathcal{N}\backslash n}} p_{n'}g_{n',k}^{t_x}(t)g_{n',k}^c(t)g_{n',k}^{r_x}(t) +N_0B},
\end{equation}
where the achievable signal-to-interference-plus-noise-ratio (SINR) term $\gamma_{k}(t)$ should, in addition to the effective receive power at vRx $k$ from vTx $n$ and to Gaussian noise, account for the effect of other interfering transmitters $n'\in\mathcal{N}\backslash n$ through their corresponding channel and antenna gains, $g_{n',k}^{c}$ and $g_{n',k}^{t_x}$, $g_{n',k}^{r_x}$ respectively. 

%%%%%%%%%%%%%%
\subsection{Evaluation of Information Value}
The contextual value gained by sharing sensing contents over an established link $(n,k)$ will be quantified in terms of: \ignore{value of the received sensing information will be quantified in terms of:}
\subsubsection{Quality/Resolution} Let assorted market penetration levels of the on-vehicle sensing equipment lead to $Q$ quality categories of vehicles with varying sensing radii $\{R_q\}_{q=1}^Q$ such that $R_q \geq R_{q'}$ if $q> q'$. Contents for a certain quality level $q\in\{1,\ldots,Q\}$ require at least $P_q^{min}$ packets successfully delivered at the destination vehicle, where $P_q^{min} \geq P_{q'}^{min}$ if $q>q'$. Therefore, the final quality level of contents received through link $(n,k)$ might be reduced due to channel conditions and the number of dropped messages during transmission. This approach models the transmission of scalable contents (e.g., multi-resolution image or scalable video coding) over V2V mmWave links.
\subsubsection{Offered Sensing Range Extension} The level of contextual novelty of the area sensed by every vTx is limited by both the urban topology of the road location (e.g., buildings without any contextual safety information for vehicles) and by the overlap between the sensed area $A_n$ by vTx $n$ and that of the receiver $k$ itself, $A_k$. Intuitively, the value of contents conveyed over link $(n,k)$ should be lower than the fraction of $A_n$ that overlaps with buildings and/or $A_k$. Circular ranges are assumed, so the normalized sensing range extension $e_{n,k}$ of vTx $n$ offered to vRx $k$ will be given by
\begin{equation}
e_{n,k} = \left(A_n \invneg A_k \invneg A_{\boxplus}\right)/\left(\pi R_{q_n}^2\right),
\end{equation}
where $q_n\in\{1,\ldots,Q\}$ is the quality of the equipment installed in vTx $n$; $A_{\boxplus}$ is the total area occupied by buildings and other urban elements lacking contextual information of interest; and the $\invneg$ operator denotes the overlaps aware area subtraction  for $e_{n,k}$ evaluation purposes.

\subsubsection{Timeliness} The interest of the content for vRx $k$ should depend on the similarity of its future route and that followed by vTx $n$ until transmission. A receiver would prefer to be matched to a transmitter if the route the latter comes from coincides with that to be followed by the former, which can be determined by e.g., comparing the programmed route at the receiver with the GPS trace of the transmitter. This similarity yields a timeliness factor $i_{n,k}\in[0,1]$ so that when delivered over link $(n,k)$, the timeliness of the content will be $\approx\hspace{-1mm}1$ if future (preceding) routes of vRx $k$ (vTx $n$) are strongly correlated.

\subsubsection{Availability}This factor stands for the delay by which packets containing the sensed content arrive at vRx $k$ through link $(n,k)$. We consider a fixed packet size $P_s$, Poisson distributed arrivals with rate $\lambda$ and a queue model for the transmitter based on a maximum buffer size of $\Delta_{\max}$ packets. When packet $p$ arrives at a queue, it is either delivered to the destination with average delay $d_{n,k}$ (given by the aggregate waiting and service time), or dropped if the entire packet is not delivered within $D_{\max}$ seconds. This queuing policy is justified by the need for low-latency communications in which new safety traffic should be prioritized over outdated contents.
%%%%%%%%%%%%%%
\subsection{Problem Formulation} \label{sec:main_problem}
The above factors can be embedded together in a single fitness function $\Psi_{n,k}(q_n,d_{n,k},e_{n,k},i_{n,k})\in[0,1]$ that evaluates the overall contextual value provided to each receiving vehicle $k$ by transmitter $n$ when paired together through mmWave link $(n,k)$. Specifically, the more valuable the connection to $n$ is for the contextual awareness of receiver $k$, the higher the value of this function. With this definition in mind, the formulation of the problem tackled in this work follows by defining a matching policy established at time $t_s\in\mathcal{T}_s$ as $\bm{\Phi}(t_s)\triangleq\{\{\phi_{n,k}(t_s)\}_{n\in\mathcal{N}(t_s)}\}_{k\in\mathcal{K}(t_s)}$ with \begin{equation}
\phi_{n,k}(t_s)=\left\lbrace
\begin{array}{ll}
\hspace{-1mm}1 &\hspace{-1mm}\text{if link $(n,k)$ is set $\forall t\in[t_s,t_s+M)$,}\\
\hspace{-1mm}0 &\hspace{-1mm}\text{otherwise,}\\
\end{array}
\right.
\end{equation}
on which we impose that a vRx $k$ (vTx $n$) should not be simultaneously matched to more than $\Omega_k\in\mathbb{N}$ vTxs\footnote{When considering concurrent transmissions from several vTxs towards the same vRx, it is implicitly presumed that each data stream is captured and processed entirely in parallel with respect to any other, without any further effect in terms of processing delay or receiving gain; the latter can be achieved by e.g., having several independent mmW transceivers deployed on each vehicle \cite{Choi2016}.} (corr. $\Omega_n\in\mathbb{N}$ vRxs) --i.e., $
\sum_{n\in\mathcal{N}(t_s)} \phi_{n,k}(t_s)\leq \Omega_k \quad \forall k\in\mathcal{K}(t_s)$--
where $\mathcal{N}(t_s)$ and $\mathcal{K}(t_s)$ denote the subset of transmitters and receivers in the scenario during the scheduling period $t_s$. To consider a subset $\mathcal{N}_k(t_s)\subseteq \mathcal{N}(t_s)$ of vTx paired to vRx $k$ a redefined fitness function $\Upsilon_{k}(\cdot)$ that extends its prior counterpart $\Psi_{n,k}(\cdot)$ is defined\footnote{For simplicity, the implications in the matching decision of a given vRx $k$ of being able to establish up to $\Omega_k$ V2V links when $\Omega_k>1$ are left for future work e.g., a mayor overlap among the areas sensed by two vTxs $\{n,m\}\in\mathcal{N}(t_s), n\neq m$ such that $e_{n,k}\approx1$, $e_{m,k}\approx1$ but $e_{n,m}\approx0$.}. Mathematically,
\begin{equation}
\Upsilon_{k}(\bm{\Phi}(t_s))\equiv\Upsilon_{k}(\{q_n,d_{n,k},e_{n,k},i_{n,k}\}_{n\in\mathcal{N}_k(t_s)})\label{eq:fitnessNoWeight}
\end{equation}
where $\Upsilon_{k}(\bm{\Phi}(t_s))\in[0,1]$ and $\mathcal{N}_k(t_s)\triangleq \{n\in\mathcal{N}(t_s): \phi_{n,k}(t_s)=1\}$. The problem addressed in this work can be formulated as the selection of the matching policy $\bm{\Phi}(t_s)$ for $t_s\in\mathcal{T}_s$ such that
\begin{subequations} \label{eq:mainOptprb}
	\begin{align}
	&\hspace{5.2mm}\underset{\bm{\Phi}(t_s)}{\text{Maximize}}&&\hspace{-2.2mm}\sum\limits_{k\in\mathcal{K}(t_s)} \Upsilon_{k}(\bm{\Phi}(t_s)), \label{eq:fitness}\\
	& \quad\;\text{subject to:} && \hspace{-2.2mm}\sum\limits_{k\in \mathcal{K}(t_s)}\phi_{n,k}(t_s)\leq \Omega_n, \forall n \in \mathcal{N}(t_s), \label{eq:matchres1}\\
	&&&\hspace{-2.2mm}\sum\limits_{n\in \mathcal{N}(t_s)}\phi_{n,k}(t_s)\leq \Omega_k,  \forall k \in \mathcal{K}(t_s), \label{eq:matchres2}\\
	&&&\hspace{-2.2mm}\phi_{n,k}(t_s)\hspace{-0.8mm} \in\hspace{-0.8mm} \{0,1\}, \forall n,k\hspace{-0.8mm}\in\hspace{-0.8mm} \mathcal{N}(t_s)\hspace{-0.8mm}\times\hspace{-0.8mm}\mathcal{K}(t_s), \label{eq:matchres3}
	\end{align}
\end{subequations}
where constraints \eqref{eq:matchres1} through \eqref{eq:matchres3} denote that every vRx can be paired to as many as $\Omega_k$ different vTxs, whereas each vTx is paired to $\Omega_n$ different vRxs at most. As opposed to the redundancy achieved by delivering sensing data through multiple transmit and receive antennas --i.e., to using multiple input multipe output (MIMO) techniques--, the above formulation considers merging sensing data coming from different sources i.e., vTxs.
%As opposed to the redundancy achieved by delivering sensing data through multiple antennas of a single vTx to a single vRx equally equipped with multiple antennas --i.e., to using MIMO techniques-- the above formulation considers merging sensing data coming from different sources i.e., vTxs. 

%%%%%%%%%%%%%%%%%%%%%%%%%%%%%%%%%%%%%%%%%%%%%%%%%%%%%%%%%%%%%%%%%%%%%%%%%%%%%%%%%%
\section{Distributed Multi-beam Matching Game} \label{sec:scheme}
%\section{Distributed Content Aware Multi-beam Association as a Matching Game}
The optimization problem in \eqref{eq:mainOptprb} is difficult to tackle analytically. Furthermore, in vehicular scenarios the design target should be steered towards low-complexity distributed solutions so as to avoid traffic overheads that could eventually compromise the end-to-end delay statistics of the deployed vehicular links. Based on this rationale we explore the framework of Matching Theory \cite{Roth1992}, \cite{Gu2015} to undertake the distributed optimization of $\bm{\Phi}(t_s)$.
\ignore{It is important to remark at this point that the ultimate purpose of this research study is to assess the performance figures of different RRM strategies, with an emphasis on \emph{reliability/delay} metrics and always considering the RRM enforcing time --- namely, the length of the scheduling interval $T_s$ --- as the driver of our analysis. In this regard, although several algorithmic alternatives from the literature will be included in the simulation benchmark later discussed in the paper, conclusions will gravitate not only on the relative performance gains among distributed association schemes, but also on the dependence of the obtained metrics with $T_s$ and its consistence over such pairing methods.}

\textbf{Definition 1:} A matching game is defined by two sets of players ($\mathcal{N}_k(t),\mathcal{K}_n(t)$) and two preference relations $\succ_n$, $\succ_k$, allowing each player $n\in {\mathcal{N}_k(t)}$, $k\in {\mathcal {K}_n(t)}$ to accordingly rank the players in the opposite set. 

The outcome of a matching game is a matching function $\bm{\Phi}(t)=\{\phi_{n,k}(t)\}$ that bilaterally assigns players $\bm{\phi}_n(t)\triangleq\{k\in \mathcal{K}_n(t):\phi_{n,k}(t)=1\}$ and $\bm{\phi}_k(t)\triangleq\{n\in \mathcal{N}_k(t):\phi_{n,k}(t)=1\}$ such that $|\bm{\phi}_k(t)|=\Omega_k$ and $|\bm{\phi}_n(t)|=\Omega_n$. Here $\Omega_n$ and $\Omega_k$ stand for the \textit{quota} of the players, vTx and vRx correspondingly. 

A preference $\succ$ is a complete, reflexive and transitive binary relation between the players in $\mathcal{N}_k(t)$ and $\mathcal{K}_n(t)$. Therefore for any vTx $n$ a preference relation $\succ_n$ is defined over the set of vRx $\mathcal{K}_n(t)$ such that for any two vRx $(k, k^\prime)\in \mathcal{K}_n(t)\times \mathcal{K}_n(t)$ with $k\neq k^\prime$, and two matchings $\bm{\Phi}(t)$ and  $\bm{\Phi}^\prime(t)$ so that $\bm{\phi}_n(t)=k$ and $\bm{\phi}_n^\prime(t)=k^\prime$:
\begin{equation}
\left(k,\bm{\Phi}(t)\right)\succ_n\left(k^\prime,\bm{\Phi}^\prime(t)\right)\Leftrightarrow U_{vTx}^{n,k}(t)>U_{vTx}^{n,k^\prime}(t).\label{eq:pref-tx}		
\end{equation}
Similarly, for any vRx $k$ a preference relation $\succ_k$ is defined over the set of vTx $\mathcal{N}_k(t)$ such that for any two vTx $(n,n^\prime)\in \mathcal{N}_k(t)\times \mathcal{N}_k(t)$ with $n\neq n^\prime$, and two matchings $\bm{\Phi}(t)$ and  $\bm{\Phi}^\prime(t)$ so that $\bm{\phi}_k(t)=n$ and $\bm{\phi}_k^\prime(t)=n^\prime$:
\begin{equation}
\left(n,\bm{\Phi}(t)\right)\succ_k\left(n\prime,\bm{\Phi}^\prime(t)\right)\Leftrightarrow U_{vRx}^{n,k}(t)>U_{vRx}^{n\prime,k}(t). \label{eq:pref-rx}		
\end{equation}
With $U_{vTx}^{n,k}(t)$ and $U_{vRx}^{n,k}(t)$ in \eqref{eq:pref-tx}-\eqref{eq:pref-rx}	respectively denoting the utility of vRx $k$ for vTx $n$ and the utility of vTx $n$ for vRx $k$ i.e., their measure of motivation of over establishing a mmWave link with a given vRx/vTx.

\textbf{Definition 2:} A matching is \emph{not stable} if for a given match $\bm{\phi}_n(t)=k$ and $\bm{\phi}_k(t)=n$, a blocking pair $(n^\prime,k^\prime)$ such that $n,n^\prime\in \mathcal{N}_k(t)$ and $k,k^\prime\in \mathcal{K}_n(t)$ satisfying $\bm{\phi}_n(t)\neq k^\prime$, $\bm{\phi}_k(t)\neq n^\prime$ and $k^\prime\succ_n j$, $n^\prime\succ_k n$ exists. That is, if for a given match two players prefer to be matched to each other over their current matched partners. A matching is considered to be \textit{pairwise stable} if no such blocking pair exists.

%%%%%%%%%%%%%%
\vspace{-1mm}
\subsection{Utilities of vRx and vTx}
Having formally described a matching game and its notion of stability, utilities for vTxs and vRxs are next defined with the twofold aim of balancing CSI/QSI information and guiding the search of prospective V2V links towards gaining more valuable traffic context information:\ignore{Utilities for vTxs and vRxs are thus formulated as follows:} 
\begin{equation} 
U_{vTx}^{n,k}\left(t_s\right)\triangleq-((R_{q_k}/R_Q)^2{\hat r}_{n,k}(t_s))^{-1},\label{eq:utx}\\
\end{equation}
\begin{equation} 
U_{vRx}^{n,k}\left(t_s\right)\triangleq-(\omega_d{\hat r}_{n,k}(t_s)+\omega_i i_{n,k}+\omega_e e_{n,k}(t_s))^{-1}. \label{eq:urx} 
\end{equation}
Moreover, $\alpha$-fair utility function \cite{Mo2000} which with $\alpha$=2 provides weighted minimum proportional delay fairness, lies at the core of the formulation of $U_{vTx}^{n,k}(t)$ as per \eqref{eq:utx} with $R_{q_k}/R_Q$ incorporating information about the resolution of vRx $k$; whereas $U_{vRx}^{n,k}(t)$ in \eqref{eq:urx} balances rate/delay and context such that $\omega_d+\omega_i+\omega_e$=1 is met. In \eqref{eq:utx}-\eqref{eq:urx}, $\hat r_{n,k}(t_s)$ represents the estimation of the average rate expected for the matched vTx $n$ and vRx $k$ pair over the next scheduling period acquired e.g., through \textit{learning} as proposed in \cite{perfecto2016millimeter}. This estimation is normalized in \eqref{eq:urx} for each vRx (corr. for each vTx in \eqref{eq:utx}).\ignore{so as to avoid \ignore{uneven} scale related issues.} 

The above matching game is solved in a distributed way using Gale-Shapley's Deferred Acceptance (DA) algorithm \cite{galeshapley1962}. DA can be applied to solve many-to-one canonical matchings i.e., those matching games where the preferences of players are not influenced by any other player's decisions, granting pairwise stability. 
\begin{figure}[t]
	\centering
	\captionsetup[subfigure]{aboveskip=1pt}
	\setlength{\textfloatsep}{2pt}
	\setlength\tabcolsep{0.5pt}
	\begin{tabular}[b]{cc}
		\multicolumn{1}{m{0.7\columnwidth}}{\multirow{2}{*}[.29\columnwidth]{
				\subfloat[\texttt{CONTEXTaware}]{\includegraphics[width=.655\columnwidth]{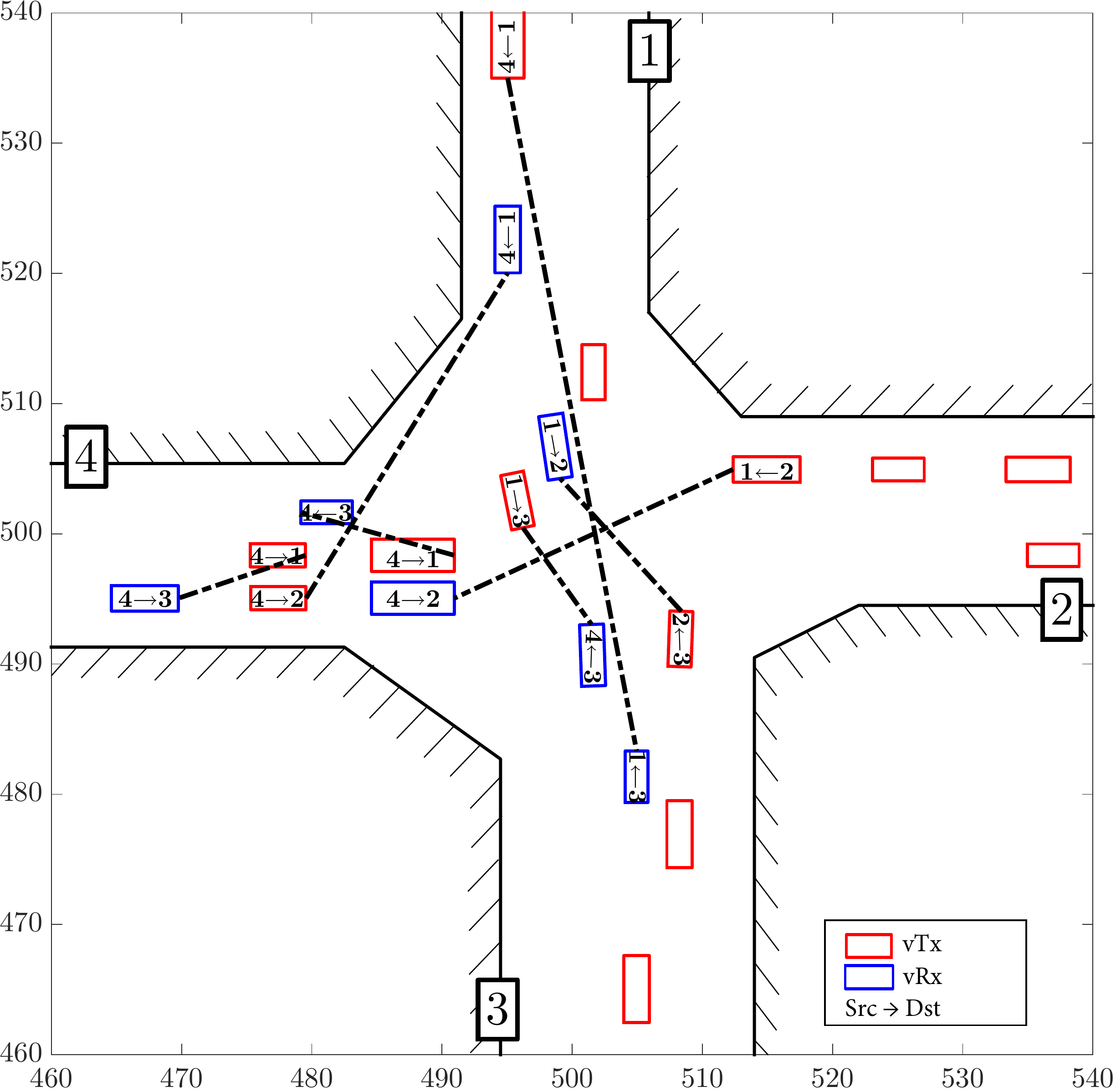}\label{fig:contextaware}}}
		}
		&\subfloat[\texttt{MINDist}]{{
				\includegraphics[width=.275\columnwidth]{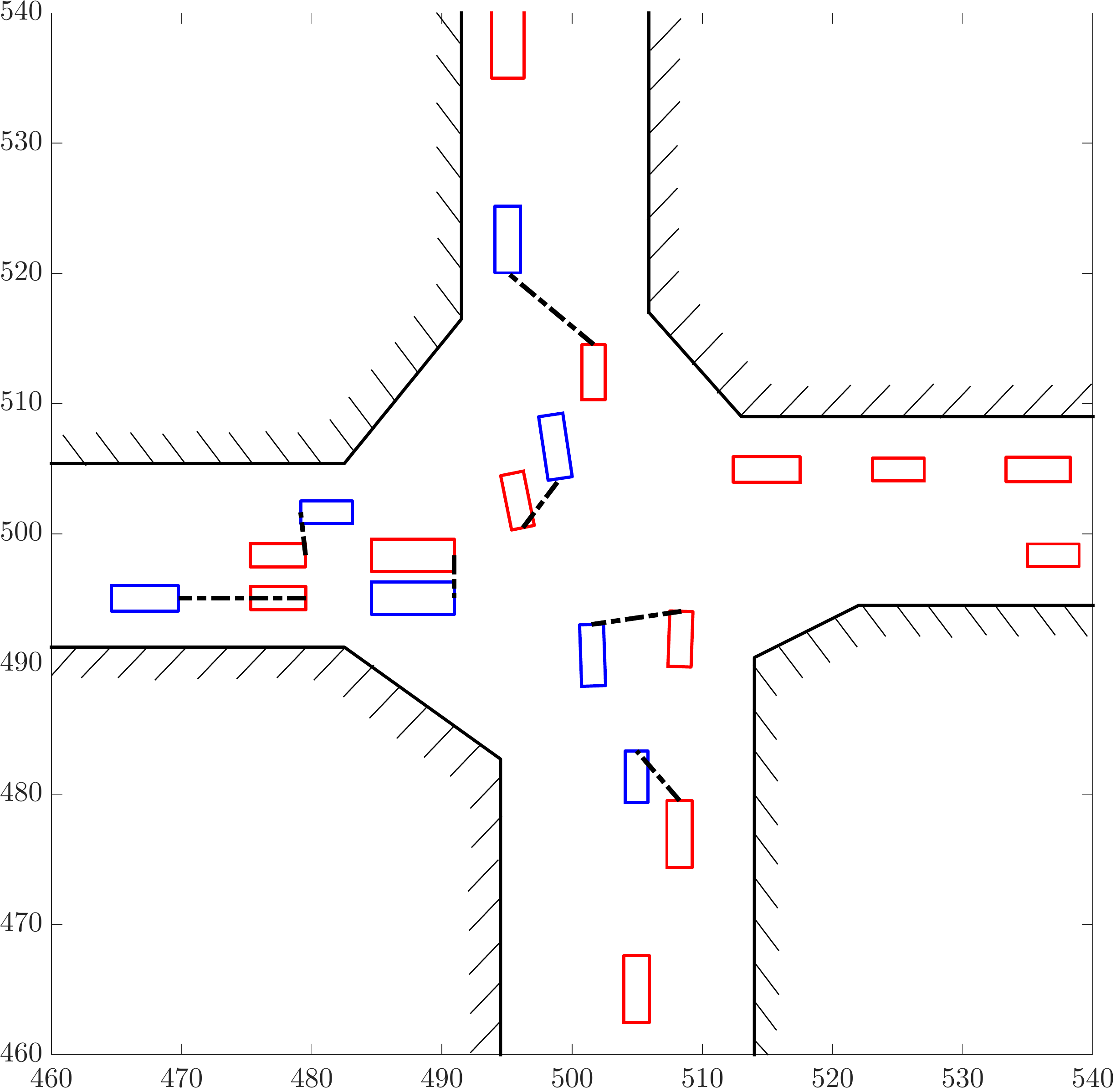}\label{fig:mindist}}}\\
		&\subfloat[\texttt{DELAYfair}]{{
				\includegraphics[width=.275\columnwidth]{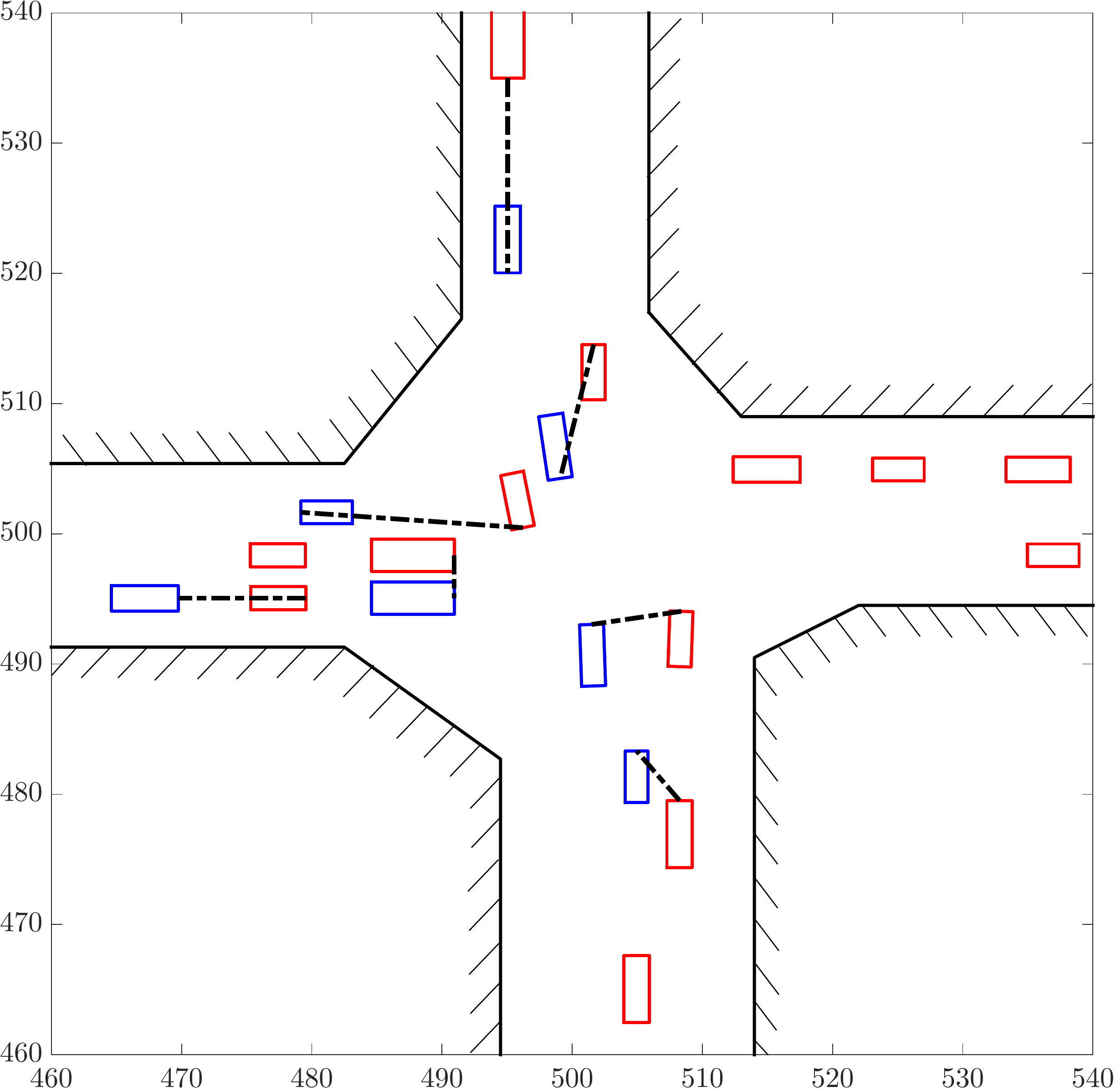}\label{fig:delayfair}}} 
	\end{tabular}		
	\caption{Influence of $\{q_n,d_{n,k},e_{n,k},i_{n,k}\}$ in the resulting V2V matching outcome (vTxs in red; vRxs in blue) for $\Omega_n=\Omega_k=1$, $t=5051$. \texttt{CONTEXTaware} with evenly balanced learned CSI/QSI and context information ($\omega_d$=0.5, $\omega_i$=0.25, $\omega_e$=0.25) is displayed in (a) with vehicles labeled `Src$\rightarrow$Dst' to ease timeliness assessment in V2V outcome. V2V link formation between closest available counterpart for \texttt{MINDist}, and based on learned CSI/QSI which includes blockage and proneness to misalignment --i.e., ($\omega_d$=1, $\omega_i$=0, $\omega_e$=0)-- for \texttt{DELAYfair} are shown in (b) and (c), respectively.}
	\label{fig:topologyresults}
\end{figure}
\setlength{\textfloatsep}{5pt}
%%%%%%%%%%%%%%%%%%%%%%%%%%%%%%%%%%%%%%%%%%%%%%%%%%%%%%%%%%%%%%%%%%%%%%%%%%%%%%%%%%
\vspace{-1mm}
\section{Simulation Setup and Numerical Results} \label{sec:results}

The performance of the proposed content-aware distributed matching scheme is assessed over a traffic light regulated junction scenario\footnote{Traffic Junction Animated \texttt{MINDist} example outcomes are available at:\\ \indent \url{https://youtu.be/2Qb4NSCbJ9w} one-to-one ($\Omega=1$)\\ \indent \url{https://youtu.be/VC3_X2fMgJA} many-to-one ($\Omega=4$) }. The scenario comprises a total of $N=26$ vTx and $K=21$ vRx of varying dimensions to emulate assorted car, bus and trucks. Vehicles' movement is characterized by realistic behavioral driving models --including acceleration, braking and lane changing-- using SUMO traces \cite{behrisch2011sumo}. Bandwidth is set to $B=2.16$ GHz centered in $60$ GHz, whereas $N_0=-174$ dBm/Hz and transmit power is equal to $p_n=15$ dBm $\forall n\in\mathcal{N}$. The simulation time spans a total of 30 seconds, slotted in transmission and scheduling periods of $T_t=2$ ms and $T_s=100$ ms ($M=50$). Transmit and receive beamwidths $\varphi_{n,k}^{t_x}$ and $\varphi_{n,k}^{r_x}$ are kept constant and equal to $\varphi\in\{5^\circ,15^\circ,45^\circ\}$ for every link. Sector-level beamwidths $\psi_n^{t_x}=\psi_k^{r_x}=45^\circ$ are adopted with $T_p/T_t=10^{-2}$. In all experiments the packet size and arrival rate are $P_s=10^6$ bits and $\lambda=1/T_t=500$ packets per second, respectively. Four different levels of sensing equipment installed on board with ranges $\{R_q\}_{q=1}^4 = \{5,10,15,20\}$ [meters] yield $4$ resolutions of the sensed contents with a minimum number $\{P_q^{min}\}_{q=1}^4=\{1,2,3,4\}$ of packets received at the destination. Queue dynamics are driven by extreme parameter values for low-latency communications, namely, a buffer size of $\Delta_{\max}=1$ packet and $D_{\max}=2$ ms.

\begin{figure}[t]
	\centering
	\captionsetup[subfigure]{aboveskip=1pt}
	\begin{tabular}{@{}l@{}}
		\subfloat[\texttt{CONTEXTaware}, $\Omega_n=\Omega_k=1$]{
			\includegraphics[width=.43\textwidth,height=.36\textwidth]{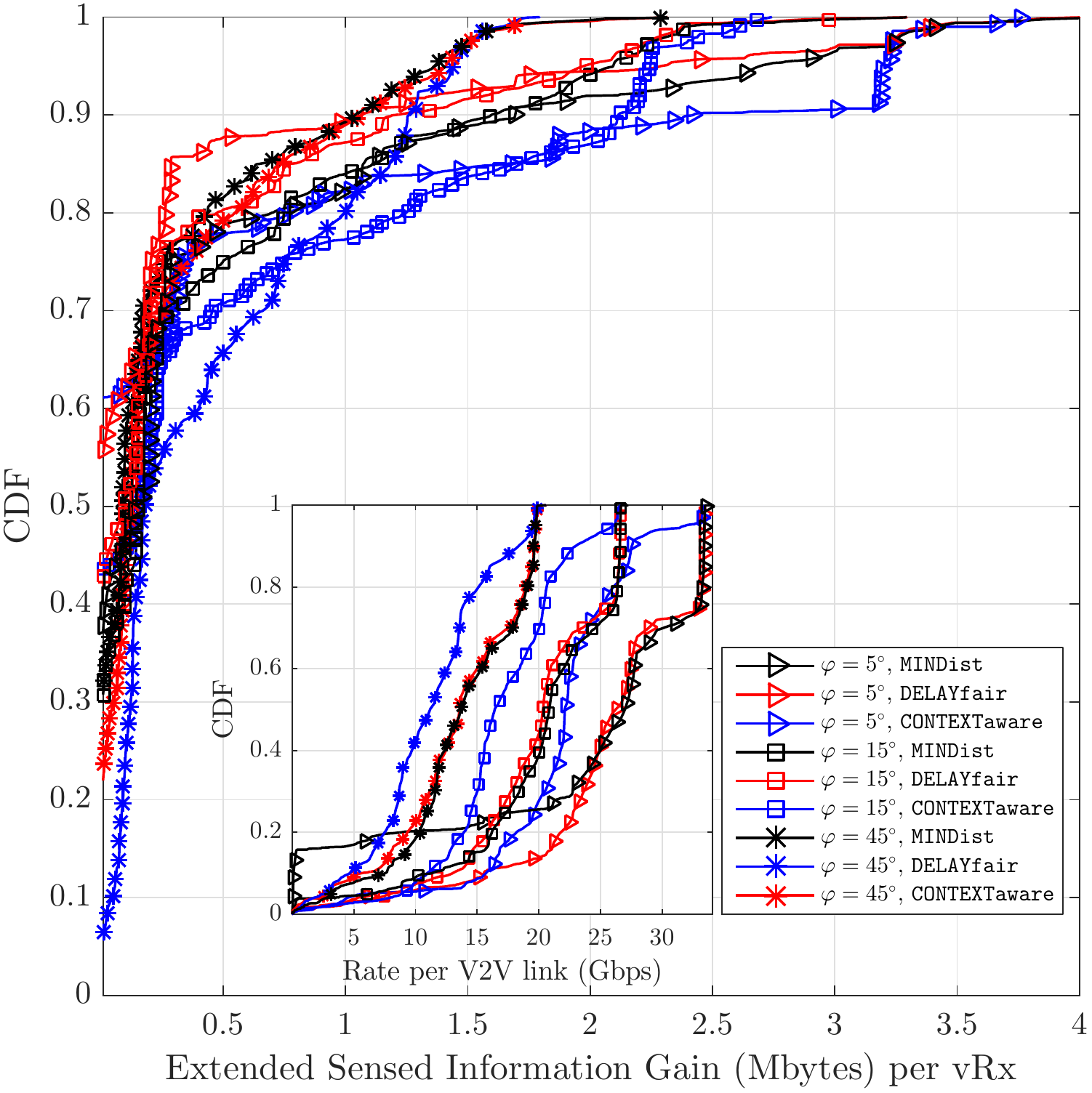}\label{fig:Info_1:1}}\\
		%		\subfloat[\texttt{CONTEXTaware}, $\Omega=2$]{
		%			\includegraphics[width=.4\textwidth]{cdf_omega_2.pdf}\label{fig:Info_2:1}}\\
		\subfloat[\texttt{CONTEXTaware}, $\Omega_n=1;\Omega_k=3$]{
			\includegraphics[width=.43\textwidth,height=.36\textwidth]{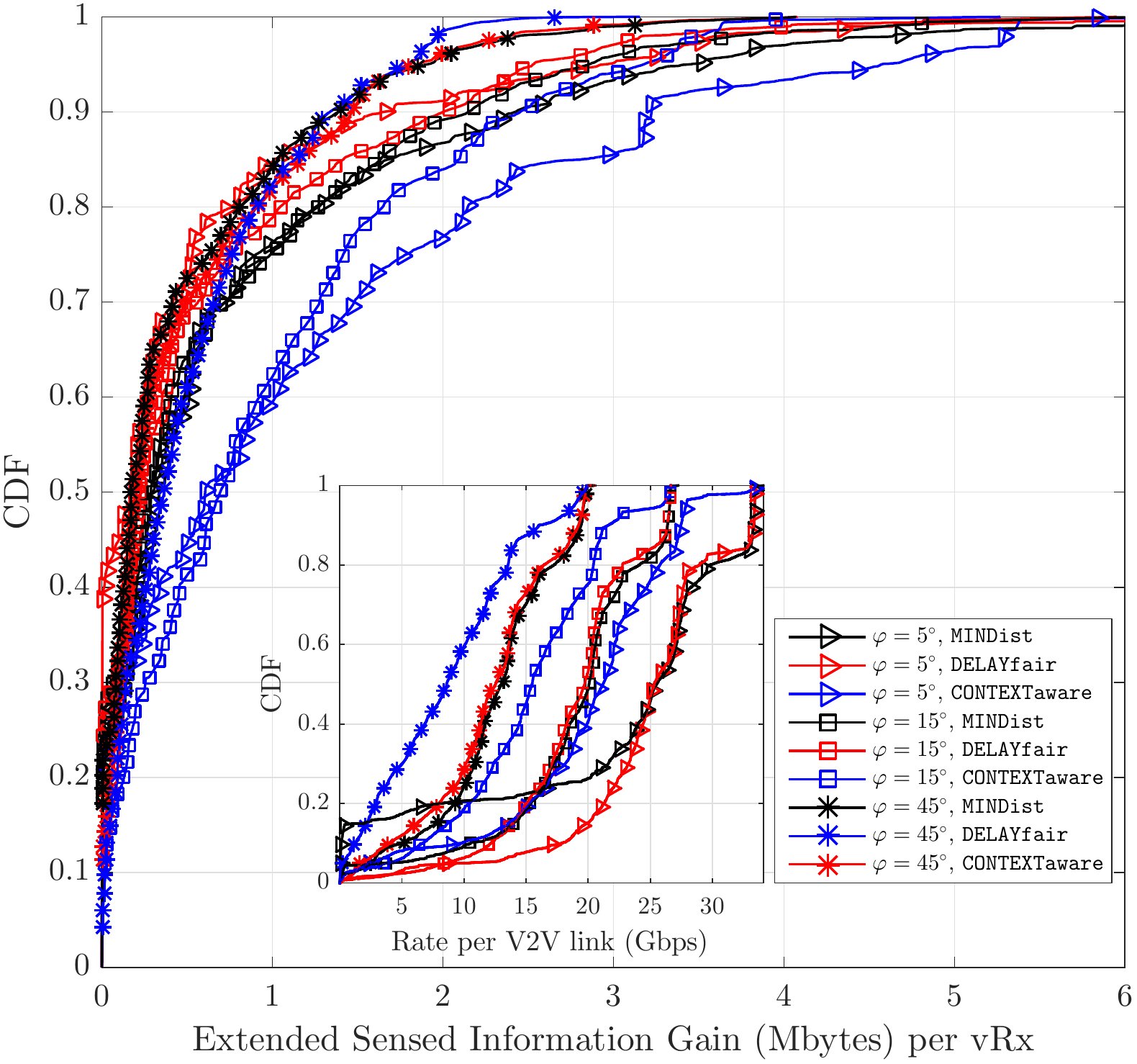}\label{fig:Info_3:1}}\\
	\end{tabular}		
	\caption{CDFs of ESI per V2V link per $T_t$. Inset CDF of Rates per V2V link.}
	\label{fig:cdfs-info-rate}
\end{figure}
The benchmark discussed in what follows considers matching outcomes \ignore{for various vTx quotas per receiver,} for three different matching policies, namely:

\begin{itemize}
\item Minimum distance matching (\texttt{MINDist}) by which vTxs and vRxs establish V2V links through a matching game that ranks players according to mutual distance.
\item Minimum-delay proportional fairness matching (\texttt{DELAYfair}), under which vTxs and vRxs are coupled \ignore{by a matching game ruled} by CSI/QSI aware weighted $\alpha$-fair utilities\ignore{(further details in \cite{perfecto2016millimeter})}.
\item Context-aware matching (\texttt{CONTEXTaware}) where the quality, novelty, timeliness and volume of the sensing content profile offered are evaluated as per \eqref{eq:utx}-\eqref{eq:urx}.
\end{itemize}
Notice that the maximum amount of feasible V2V links is subject to limitations arising from the skewness in the distribution of vTx and vRx, as well as from the impracticality of pairing vehicles if there are buildings blocking their link.\ignore{Indeed, schemes that build upon \textit{learned} CSI/QSI will never match vTx and vRx whose LOS is obstructed by a building even if that means remaining unpaired.}

\begin{figure}[t]
	\centering
	\captionsetup[subfigure]{aboveskip=1pt}
	\begin{tabular}{@{}l@{}}
		\subfloat[\texttt{CONTEXTaware}, $\Omega_n=\Omega_k=1$]{
			\includegraphics[width=.43\textwidth, height=.36\textwidth]{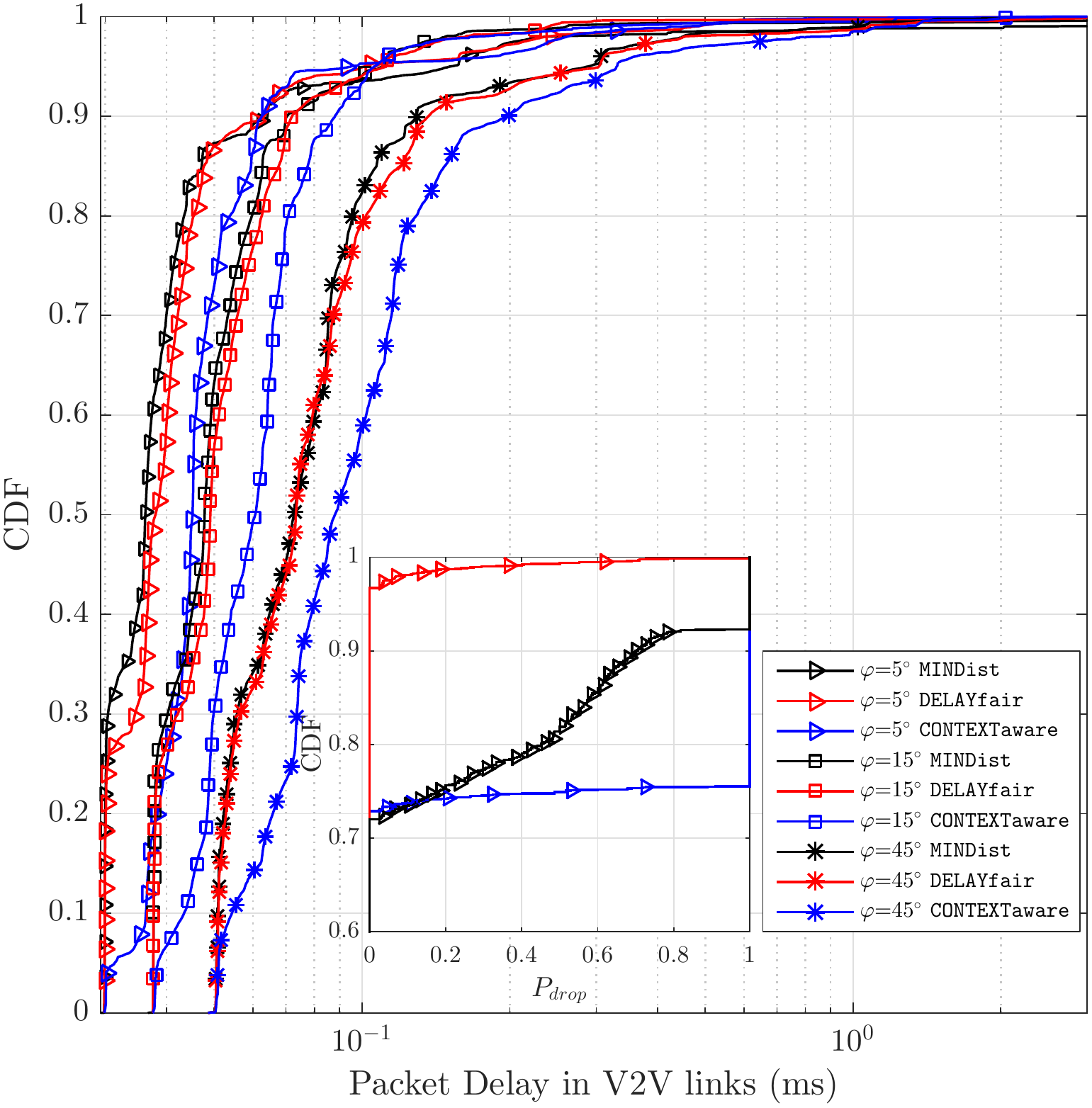}\label{fig:Delay_drop_1:1}}\\
		%		\subfloat[\texttt{CONTEXTaware}, $\Omega=2$]{
		%			\includegraphics[width=.4\textwidth]{cdf_omega_2_Drop_DelayCDF_FixedPcktSize_1000000_MeanIntArrival_1}\label{Delay_drop_2:1}}\\
		\subfloat[\texttt{CONTEXTaware}, $\Omega_n=1;\Omega_k=3$]{
			\includegraphics[width=.43\textwidth, height=.36\textwidth]{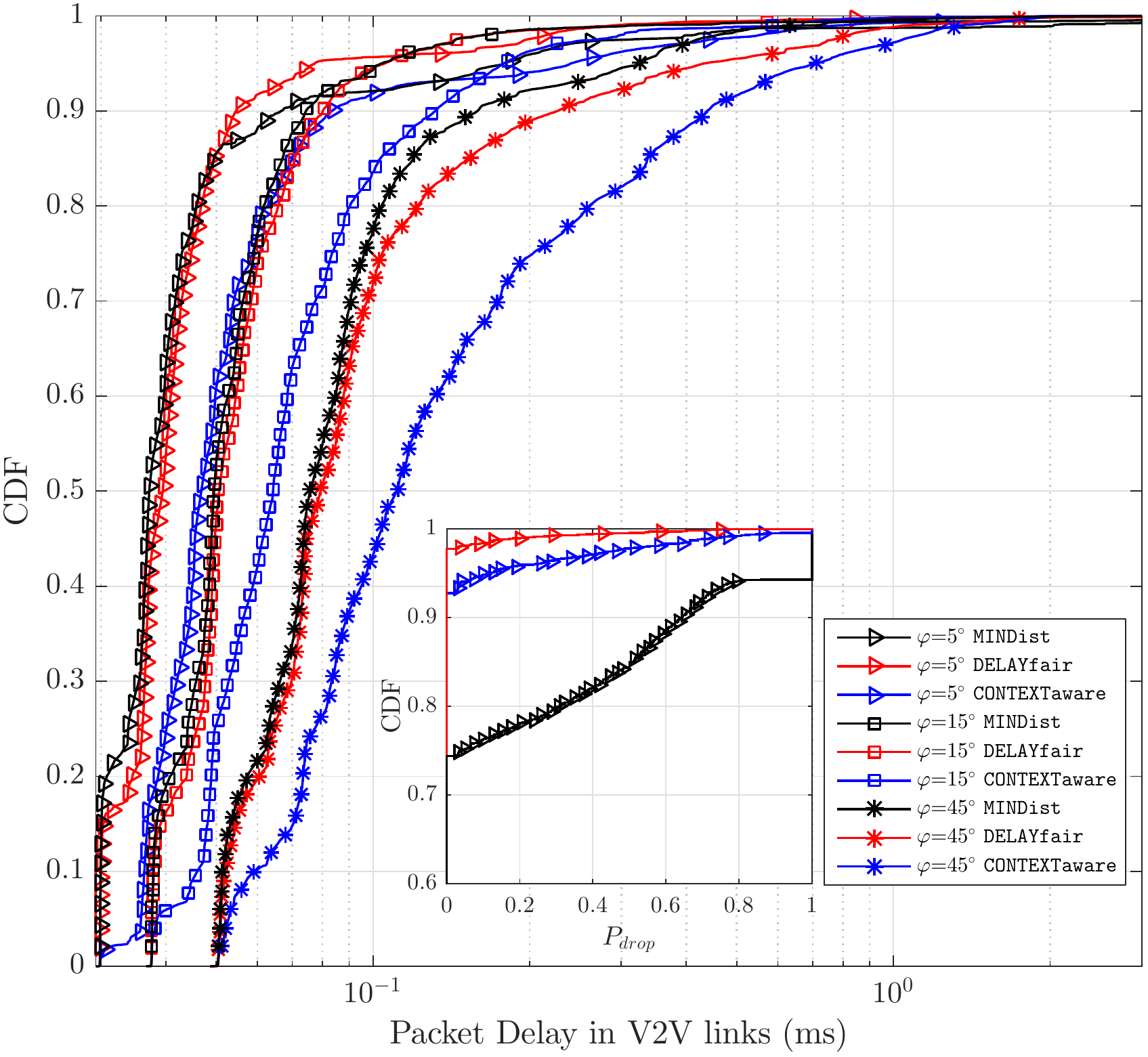}\label{fig:Delay_drop_3:1}}\\
	\end{tabular}		
	\caption{CDFs of Delay per V2V link per $T_t$. Inset CDF of $P_{drop}$ for $\varphi=5^\circ$.}
	\label{fig:cdfs-delay-drops}
\end{figure}
%%%%%%%%%%%%%%
\subsection{Discussion}
%where the matched vTx-vRx pairs for the different schemes included in the benchmark are depicted for one of the simulated scheduling intervals and $\Omega=1$.
We will first focus the analysis of the results on Fig.~\ref{fig:topologyresults}; the aim of these plots is to evince the topological differences yielded by the diverse matching criteria of the policies included in the benchmark. To begin with, \texttt{MINDist} \ignore{pairs vehicles exclusively based on distance which} gives rise to short-length mmWave links \ignore{prone to frequent misalignments} as depicted by Fig.~\ref{fig:topologyresults}(b). \ignore{\ref{fig:mindist}}The adoption of learning techniques in the utility formulation for \texttt{DELAYfair} paves the way for the discovery of better long-term aligned pairs and, as can be observed in Fig.~\ref{fig:topologyresults}(c)\ignore{\ref{fig:delayfair}}, leads to topological changes in the V2V link selection with respect to those of \texttt{MINDist}. Finally, proposed \texttt{CONTEXTaware} approach with $\omega_d=0.5$ and $\omega_i=\omega_e=0.25$, balances learning results from link level performance metrics and the potential amount of contextual novelty and timeliness provided by vTxs to every receiver as per \eqref{eq:utx}-\eqref{eq:urx}. As shown in Fig.~\ref{fig:topologyresults}(a)\ignore{\ref{fig:contextaware}}, mmWave links  \ignore{between vehicles} under this scheme are enforced not only accounting for their delay statistics, but also by the contextual information gain, as quantified by minimal overlaps between sensing areas and timeliness --expressed through `src$\rightarrow$dst' labels--, of the information with respect to the vehicles' routes. 

We follow with a discussion of Fig.~\ref{fig:cdfs-info-rate} and Fig.~\ref{fig:cdfs-delay-drops}, where the cumulative distribution functions (CDF) of 1) the extended sensed information (ESI) gained per vRx and transmission slot; and 2) the packet level experienced delay are depicted for $\Omega_k=\{1,\ignore{2,}3\}$. Fig.~\ref{fig:cdfs-info-rate}(a) and Fig.~\ref{fig:cdfs-info-rate}(b) \ignore{Fig.~\ref{fig:Info_1:1} and Fig.~\ref{fig:Info_3:1}} include the CDFs of the instantaneous rates per V2V link as subplots, whereas Fig.~\ref{fig:cdfs-delay-drops}(a) and Fig.~\ref{fig:cdfs-delay-drops}(b) \ignore{Fig.~\ref{fig:Delay_drop_1:1} and Fig.~\ref{fig:Delay_drop_3:1}} include CDFs of packet drop ($P_{drop}$) --obtained by aggregating successful and failed transmission over each scheduling period and queue-- to illustrate the different trade-offs of the simulated scenario. The ESI in bits is defined as 
\begin{equation} 
\mbox{ESI}_{k}(t)=\hspace{-3mm}\sum_{n\in \mathcal{N}_k(t)}\phi_{n,k}(t_s) i_{n,k} e_{n,k}(R_{q_n}/R_Q)^2r_{n,k}(t) T_t,
\end{equation}
where $t\in [t_s,t_s+M)$. This notion of ESI reflects the maximum information the rate of an established mmWave link supports to deliver within a transmission slot $T_t$ weighted by the timeliness and the sensing resolution and range extension area brought by the vTx. A quick glimpse at Fig.~\ref{fig:cdfs-info-rate}(a) \ignore{Fig.~\ref{fig:Info_1:1}} exposes the superior performance of the proposed scheme providing 43\% (percentile 90, $\varphi=5^\circ$) and 67\% (percentile 80, $\varphi=15^\circ$) more information than the best of the other baselines no matter the comparatively lower rates achieved in V2V links. The same result holds for Fig.~\ref{fig:cdfs-info-rate}(b) \ignore{Fig. \ref{fig:Info_3:1}} with 33\% (percentile 90) and 71\% (percentile 80) increased information in both cases for $\varphi=5^\circ$. As for results in Fig.~\ref{fig:cdfs-delay-drops}(a), \ignore{Fig. \ref{fig:Delay_drop_1:1}} values of $\varphi=\{5^\circ,15^\circ\}$ for all schemes show delays under $0.1$ ms for over 90\% of the samples (80\% for $\varphi=45^\circ$ in baselines, 60\% in our proposed matching). Similarly, our proposed approach shows a slightly worse delay performance in Fig.~\ref{fig:cdfs-delay-drops}(b) \ignore{Fig. \ref{fig:Delay_drop_3:1}} that is directly related to a significantly improved reliability (93\% of successful transmissions) when vRxs are allowed to leverage information arriving from assorted vTxs. The reason lies in the many-to-one matching, which increases the chances of vTxs to match vRxs on top of their ranks i.e., those offering better CSI/QSI profiles; a better CSI/QSI implies reduced drops and more samples contributing to delay calculations. The penalty to be paid, however, comes in the form of a few vRxs monopolizing the access to multi-sourced shared sensing information.    

%%%%%%%%%%%%%%%%%%%%%%%%%%%%%%%%%%%%%%%%%%%%%%%%%%%%%%%%%%%%%%%%%%%%%%%%%%%%%%%%%%
\section{Conclusions and Future Research Lines} \label{sec:conclusions}
%Framed on the ever-growing need for a higher contextual awareness in vehicular environment,
This work has elaborated on how to share contextual sensed information among vehicles using many-to-one mmWave V2V links. In particular, a distributed matching game based on a joint delay and information value utility formulation has been proposed. Simulation results have been discussed for a road junction scenario with realistic mobility traces, from where it is concluded that our proposed scheme outperforms in terms of ESI baselines that only consider link level metrics, with a toll on slightly increased delays and higher drops. These results highlight the need to incorporate the contextual value of the information conveyed through V2V mmWave links into the pairing strategies to effectively extend vehicles' individual road/traffic awareness. Future research efforts will be aimed at designing more realistic models based on trajectories for both information timeliness and offered sensing extension, and to their application in collaborative dynamic map building.
%sensing extension based on trajectories. Their application to collaborative dynamic map building will be also actively investigated.
%sensing extension based on trajectories and prediction to further applying them to collaborative dynamic map building.

%Simulations have been presented and discussed for a road junction scenario with realistic mobility traces, from where it is concluded that the set of formulated utilities permit to pair vehicles combining different performance objectives: rate/delay, the extended sensing area provided by the transmitter upon pairing and the timeliness of the exchanged information with respect to the vehicles' routes.

%%%%%%%%%%%%%%%%%%%%%%%%%%%%%%%%%%%%%%%%%%%%%%%%%%%%%%%%%%%%%%%%%%%%%%%%%%%%%%%%%%
\section{Acknowledgments} \label{sec:ack}
This research was supported by the Basque Government under ELKARTEK program (BID3ABI project) and by the Spanish Ministerio de Economia y Competitividad (MINECO) under grant TEC2016-80090-C2-2-R (5RANVIR).
\vspace{-1mm}
\bibliographystyle{IEEEtran}
\bibliography{IEEEabrv,EuCNC}
\end{document}